\documentclass[prd,aps,twocolumn,showpacs,draft]{revtex4}

\begin{document}

\title{General Solution for the Static, Spherical and Asymptotically Flat 
Braneworld\footnote
{This paper is based on a note submitted to the Japanese Physical Society 2011 Annual Meeting which was canceled due to the earthquake in Tohoku district in March.
}
}
\author{			Keiichi Akama}
\author{			Takashi Hattori$^{\rm A}$}
\author{			Hisamitsu Mukaida}
\affiliation{	Department of Physics, Saitama Medical University,
 			 Saitama, 350-0495, Japan}
\affiliation{	$^{\rm A}$Department of Physics, Kanagawa Dental College,
 			 Yokosuka, 238-8580, Japan}
\date{\today}

\begin{abstract}
The general solution for the static, spherical and asymptotically flat braneworld
	is derived by solving the bulk Einstein equation and braneworld dynamics. 
We show that it involves a large arbitrariness,
	which reduces the predictive powers of the theory.
Ways out of the difficulty are discussed. 
\end{abstract}




\maketitle

Einstein gravity is successful in explaining 
	(i) the Newton's law of universal gravitation 
	for moderate distances, and 
	(ii) the post-Newtonian evidences (light deflections and
	planetary perihelion precessions due to solar gravity, etc.).
The explanations are achieved via the Schwarzschild solution 
	of the Einstein equation
	based on the ansatze 
	(a) staticity, 
	(b) spherical symmetry,
	(c) asymptotic flatness, and 
	(d) emptiness.
On the other hand, there are many plausible reasons that tempt us
	to take our 3+1 dimensional curved spacetime 
	as a ``braneworld" embedded in higher dimensions (``bulk")
	\cite{Fronsdal}--\cite{Jardim:2011gg}.
There exist naive expectations that
	the braneworld theories inherit the successes of the Einstein gravity. 
It is, however, not a trivial problem because 
	they are not based on the brane Einstein equation, 
	but on the coupled system of the bulk Einstein equation and 
	the equation of motion of the braneworld.
In the previous paper, we obtained the general solution
	of the diagonal components of the equations at the brane
	\cite{Akama:2010rb}.
In this paper, we derive the general solution
	of the full set of the fundamental equations all over the bulk
	under the {Schwarzschild ansatz} (a)--(d).
We find that it involves a large arbitrariness,
	which reduces the predictive powers of the theory.
For the braneworlds, the {Schwarzschild ansatze} require
	more precise specifications of the meaning.
We impose the staticity (a) all over the bulk, 
	while the asymptotic flatness (c) only on the braneworld,
	and not necessarily outside.
As for the sphericity (b), we assume symmetry under all the global rotations 
	which maps the braneworld onto itself.
We assume the emptiness (d) only outside the braneworld, 
	and allow matter and energy distributions 
	within a thin region inside the brane.
The empty region is not necessarily infinite, 
	but can be bounded due to topological structures or 
	other objects.

In order to seek for the general solution,
	we begin with examining what are the equations to be solved.
For definiteness, we consider the 3+1 dimensional Nambu-Goto type brane 
	in 4+1 curved spacetime with Einstein type gravity.
Let $X^I$ be the bulk coordinate, and
	$g_{IJ}(X^K)$ be the bulk metric 
	at the point $X^K$ \cite{notation}.	
Let the brane be located at $X^I=Y^I(x^\mu)$ in the bulk, 
	where $x^\mu$ ($\mu=0,1,2,3$) are parameters
	which serve as the brane coordinate.
The dynamical variables of the system are 
	$g_{IJ}(X^K)$, $Y^I(x^\mu)$, and matter fields. 
The $Y^I(x^\mu)$ should be taken as the collective modes of the brane 
	formed by matter interactions.
Note that the brane induced metric 
	$\tilde g_{\mu\nu}=Y^I_{,\mu} Y^J_{,\nu} g _{IJ}(Y^K)$
	of the brane cannot be a dynamical variable,
	since it alone cannot completely specify the state of the braneworld
	\cite{notation}.
Then, the action integral is given by
\begin{eqnarray}&&
	S=S_g+S_Y+ S_{\rm m}\ \ \ {\rm with}
  \label{action}
\\&&
	S_g=\int\!\!\sqrt{- g}( -2\lambda+ \kappa^{-1} R)d^5 X,\ \ \ \ 
\\&&
	S_Y=-2\tilde\lambda \!\int\!\! \sqrt{-\tilde g } d^4 x,\ \ \ \ 
\end{eqnarray}
where 	$ S_{\rm m}$ is the matter action,
	$g=\det g_{IJ}$, 
	$ R = R^I_{\ I}$,
	$ R_{IJ}= R^K_{\ IJK}$, 
	$ R^L_{\ IJK}$ 
	is the bulk curvature tensor
	written in terms of $ g _{IJ}$,
	$\tilde g =\det \tilde g_{\mu\nu} $, 
	and  $ \lambda $, $ \kappa$, and $\tilde \lambda$ 
	are constants.
We add no artificial fine-tuning terms.
The equations of motion derived from (\ref{action}) are %
\begin{eqnarray}&&
	E_{IJ}\equiv R_{IJ}-\frac{1}{2} R g _{IJ}
	+ \kappa (T_{IJ}+ \lambda g _{IJ})=0,
  \label{BE}
\\&&
	 (\tilde \lambda \tilde g^{\mu\nu}
	+\tilde T^{\mu\nu})
	Y^I_{;\mu\nu}=0,\ \ \ \ \ \ 
  \label{NG}
\end{eqnarray}
and those for matters,
where 	
	$T_{IJ}$ and $\tilde T_{\mu\nu}$ are the energy momentum tensors 
	with respect to $g_{IJ}$ and $\tilde g_{\mu\nu}$, 
	respectively \cite{notation}.
Eq.\ (\ref{BE}) is the bulk Einstein equation, and 
	eq.\ (\ref{NG}) is the Nambu-Goto equation
	for the {braneworld dynamics}. 
The Einstein equation of the brane does not exist.

In accordance with the ansatze of staticity and sphericity, 
	we introduce time, redial, polar and azimuth coordinates,
	$X^0=t$, $X^1=r$, $X^2=\theta$, and $ X^3=\varphi$, respectively,
	and the normal geodesic coordinate $X^4=z$, 
	such that $X^\mu=x^\mu$ ($\mu=0,\cdots,3$) and $z=0$ at the brane.
According to staticity and sphericity we can generally choose 
	the metric tenser $ g _{IJ}$ of the form 
\begin{eqnarray}&&
	ds^2= g _{IJ} dX^I dX^J
\cr&&
	=fdt^2-hdr^2-k(d\theta^2+\sin^2\theta d\varphi^2)-dz^2,
  \label{ds2}
\end{eqnarray}
where	$f$, $h$ and $k$ are functions of $r$ and $z$ only,
	and we choose as $k|_{z=0}=r^2$ using diffeomorphism.
Asymptotic flatness of the brane implies that 
\begin{eqnarray}&&
	f,\ h\rightarrow 1\ \ \ \
	{\rm as} \ \ \ \ r\rightarrow\infty\ \ \ \  
	{\rm at} \ \ \ \ z=0. 
  \label{f,h->1}
\end{eqnarray}
The independent non-vanishing components of the Ricci tensor $R_{IJ}$ are
\begin{eqnarray}&&
	R_{00}=-f_{zz}/2+ f_z{}^2/4f - f_z h_z/4h- f_z k_z/2k
\cr&&\ \ \ \ \ \ \ 
	- f_{rr}/2h+ f_r{}^2/4fh+ f_r h_r/4h^2- f_r k_r/2kh,\ \ \ \ \ 
\label{R00}
\\&&%
	R_{11}=h_{zz}/2-h_z{}^2/4h
	+{f_{z} h_{z}}/{4f}+ {h_{z} k_{z}}/{2k}
\cr&&\ \ \ \ \ \ \ \ \ 
	+ f_{rr}/2f- f_r{}^2/4f^2- f_r h_r/4fh\ \ \ \ 
\cr&&\ \ \ \ \ \ \ \ \ \ 
	+ k_{rr}/k- k_r{}^2/2k^2- k_r h_r/2kh,\ \ \ \ 
\label{R11}
\\&&%
	R_{22}= k_{zz}/2+ f_z k_z/4f+ h_z k_z/4h
\cr&&\ \ \ \ \ \ \ \ \ 
	+ k_{rr}/2h+f_r k_r/4fh- h_r k_r/4h^2-1, \ \ \ \ 
\label{R22}
\\&&%
	R_{44}= f_{zz}/2f+ h_{zz}/2h+ k_{zz}/k
\cr&&\ \ \ \ \ \ \ \ \ 
	- f_z{}^2/4f^2- h_z{}^2/4h^2- k_z{}^2/2k^2, \ \ \ \ 
\label{R44}
\\&&%
	R_{14}= f_{zr}/2f + k_{zr}/k - f_z f_r/4f^2
\cr&&\ \ \ \ \ \ \ \ \ 
	- h_z f_r/4fh- h_z k_r/2hk - k_z k_r/2k^2, \ \ \ \ 
\label{R14}
\end{eqnarray}
where subscripts $r$ and $z$ indicate partial differentiations.

We first solve the bulk Einstein equation (\ref{BE}) alone
	without brane dynamics (\ref{NG}).
For a while, we also discard emptiness assuming 
	that $T_{IJ}$ is not necessarily vanishing.
Staticity and sphericity imply that only $T_{00}$, $T_{11}$, $T_{22}$, 
	$T_{44}$ and $T_{14}$ among $ T_{IJ}$ are independent, 
	and functions of $z$ and $r$ only.
We rewrite the equation (\ref{BE}) into the equivalent form
\begin{eqnarray}
	{{F}}_{IJ} \equiv E_{IJ} - E^K_ {\ K } g _{IJ}/3 =0, 
  \label{FIJ}
\end{eqnarray}
where  	those with $(I,J)$=(0,0), (1,1), (2,2), (4,4) and (1,4)
	are independent.
We have five equations for three functions $f$, $h$ and $k$.
Now we show that the five equations are not all independent.
The Bianchi identity of the curvature tensor
	and covariant conservation low of energy-momentum tensor imply
\begin{eqnarray}&&\hskip-15pt
	{{F}}_{14,4} = -{{F}}_{00,1}/2f
	- {{F}}_{11,1}/2h+{{F}}_{22,1}/k
	+{{F}}_{44,1}/2
\cr&&
	-{{F}}_{11}\{\ln(k^2f/h)\}_{,1}/2h
	-{{F}}_{14}\{\ln(k^2fh)\}_{,4}/2, 
  \label{F14,4}
\\&& \hskip-15pt
	{{F}}_{44,4} =-{{F}}_{00,4}/f
	+ {{F}}_{11,4}/h+2{{F}}_{22,1}/k
	- 2{{F}}_{14,1}/h
\cr&&
	-{{F}}_{14}\{\ln(k^2f/h)\}_{,1}/h
	-{{F}}_{44}\{\ln(k^2fh)\}_{,4}. 
  \label{F44,4}
\end{eqnarray}
We expand every quantity in terms of $z$. 
We denote by $A^{[n]}(r)$ the coefficient of the $z^n$ term 
	in the expansion of any function $A(r,z)$:
\begin{eqnarray}
	A(r,z)=\sum_{n=0}^{\infty}A^{[n]}(r) z^n.
\end{eqnarray}
Let us assume ${{F}}_{00}= {{F}}_{11}= {{F}}_{22}=0$.
Then, we have	
\begin{eqnarray}&& \hskip-10pt
	{{F}}_{14}^{[n]} =
	[{{F}}_{44,1}/2-{{F}}_{14}\{\ln(k^2fh)\}_{,4}/2] ^{[n-1]}/n,
  \label{F14n}
\\&& \hskip-10pt
	{{F}}_{44}^{[n]} =
	[- 2{{F}}_{14,1}/h-{{F}}_{14}\{\ln(k^2f/h)\}_{,1}/h 
\cr&& \ \ \ \ \ \ \ \ \ \ \ \ \ \ \ \ \ \ \ \ \  
	-{{F}}_{44}\{\ln(k^2fh)\}_{,4}	]^{[n-1]} /n
  \label{F44n}
\end{eqnarray}
for $n\ge1$.
The right-hand side of eqs. (\ref{F14n}) and (\ref{F44n})
	are linear combinations of ${{F}}_{14}^{[j]}$ and ${{F}}_{44}^{[j]}$
	with $0\le j \le n-1$ and their $r$-derivatives.
Therefore, if we have ${{F}}_{14}^{[0]}= {{F}}_{44}^{[0]}=0$, 
	we can conclude that ${{F}}_{14}^{[n]}= {{F}}_{44}^{[n]}=0$
	for any $n\ge1$.
Thus, the independent equations to be solved are 
\begin{eqnarray}&&
	{{F}}_{00}= {{F}}_{11}= {{F}}_{22}=0,       \label{F00=F11=F22=0}
\\&& 
	{{F}}_{14}^{[0]}= {{F}}_{44}^{[0]}=0.   \label{F140=F440=0}
\end{eqnarray}
From the viewpoints of (\ref{F00=F11=F22=0}) and (\ref{F140=F440=0})
	the treatment in Ref.\ \cite{Akama:2010rb} is insufficient, 
	because there we considered only
	${{F}}_{00}= {{F}}_{11}= {{F}}_{22}={{F}}_{44} =0$
	on the brane.
It was, however, reasonable that we considered only on-brane equations,
	since only they are essential to determine the brane physics, 
	as will be seen soon.
The equations in (\ref{F00=F11=F22=0}) imply the recursion formulae
\begin{eqnarray}&&
	f^{[n]}= {1 \over n(n-1)}\bigg[{f_z{}^2 \over 2f} - {f_z h_z\over 2h}
	- {f_z k_z \over k}- {f_{rr} \over h}+ {f_r{}^2 \over 2fh}
\cr&&\ 
	+ {f_r h_r \over 2h^2}- {f_r k_r \over kh}
	+ 2\kappa \left(T_{00} - {T f \over3} - {2\lambda f\over3}\right)
	\bigg]^{[n-2]},\ \ \ \ \ 
\label{fn}
\\&&%
	h^{[n]} ={1 \over n(n-1)}\bigg[{ h_z{}^2 \over 2h}
	-{f_{z} h_{z} \over {2f}}- {h_{z} k_{z} \over {k}}
\cr&&\ \ \ 
	- {f_{rr} \over f}+ {f_r{}^2 \over 2f^2}+ {f_r h_r \over 2fh}
	- {2k_{rr} \over k}+ {k_r{}^2 \over k^2}+ {f_r k_r \over kh} 
\cr&&\ \ \ \ \ \ \ 
	- 2\kappa \left(T_{11} + {T h \over3} + {2\lambda h\over3}\right)
	\bigg]^{[n-2]},\ \ \ \ \ 
\label{hn}
\\&&%
	k^{[n]}= {1 \over n(n-1)}\bigg[- {f_z k_z\over 2f}- {h_z k_z\over h}
	-{ k_{rr}\over h}+ {f_r k_r\over 2fh}
\cr&&\ \ \ \ 
	+ {h_r k_r\over 2h^2}+2
	- 2\kappa \left(T_{22} + {T k \over3} + {2\lambda k\over3}\right)
	\bigg]^{[n-2]} \ \ \ \ \ 
\label{kn}
\end{eqnarray}
for $n\ge2$.
The right-hand side of eqs.\ (\ref{fn})--(\ref{kn})
	are written in terms of
	$f^{[j]}$ $h^{[j]}$, and $k^{[j]}$ for $0\le j\le n-1$.
This means that, if we have $f^{[0]}$ $h^{[0]}$, $k^{[0]}$,
	$f^{[1]}$ $h^{[1]}$, and $k^{[1]}$, 
	we can recursively determine $f^{[n]}$ $h^{[n]}$, and $k^{[n]}$
	for any $n\ge2$.
Therefore, we can uniquely derive $f$, $h$ and $k$ 
	in a bulk region around the brane 
	if we know $f$, $h$, $k$, $f_z$, $h_z$ and $k_z$ on the brane.
They are, however, not all independent, 
	but should obey eq.\ (\ref{F140=F440=0}).
For notational simplicity, we define the variables 
	$\phi$, $\psi$, $a$, $b$ and $c$ with
\begin{eqnarray}&&
	f^{[0]}=e^{2\phi},\ \ 
	h^{[0]}= e^{2\psi},\ \  
\\&&
	f^{[1]} =-2a f^{[0]},\ \ 
	h^{[1]} =-2b h^{[0]},\ \ 
	k^{[1]} =-2c k^{[0]}.\ \ \ \ 
\end{eqnarray}
Remember that we have chosen as $k^{[0]}=r^2$ using diffeomorphism.
The matrix diag($a,b,c,c$) gives the extrinsic curvature of the brane.
Eliminating $f^{[2]}$, $h^{[2]}$ and $k^{[2]}$ from eq.\ (\ref{F140=F440=0})
	with (\ref{fn})--(\ref{kn}) for $n=2$, we obtain
\begin{eqnarray}&&
	a_r+2c_r+(a-b)\phi_r+2(c-b)/r=\kappa \tau_{14},
       \label{F140=0}
\\&& 
	e^{-2\psi}[\phi_{rr} + (\phi_r -\psi_r)( \phi_r+2/r)
	-( e^{2\psi}-1)/r^2] \ \ \ \ \ 
\cr&&\ \ \ \ \ \ \ \ \ 
	+ab+2ac+2bc+c^2=\kappa (\tau_{44}-\lambda), 
\label{F440=0}
\end{eqnarray}
where $\tau_{IJ}=T_{IJ}|_{z=0}$.
The first line of (\ref{F440=0}) is equal to $ -\tilde R/2$,
	where $\tilde R$ is the scalar curvature of the brane.
Note that $\tilde R$ is not equal to the bulk scalar curvature 
	$R$ at the brane.

Thus, we have two equations (\ref{F140=0}) and (\ref{F440=0})
	for five functions $\phi$, $\psi$, $a$, $b$ and $c$
	for a given set of $\tau_{14}$ and $\tau_{44}$.
Therefore, the solution includes three arbitrary functions.
Now, we choose $a$, $b$ and $c$ arbitrarily.
If $a\not=b$, 
	eqs.\ (\ref{F140=0}) and (\ref{F440=0}) are solved 
	to give the solution:
\begin{eqnarray}&&
	\phi=-\int_r^\infty A dr,\ \ \ 
       \label{phi=}
\\&& 
	2\psi=-\int_r^\infty {P}dr
	-\ln\left(\eta-\int_r^\infty Qe^{-\int_r^\infty Pdr}dr
	\right)\ \ \ \ ,
\label{psi=}
\end{eqnarray}
with an arbitrary constant $\eta$ and the functions
\begin{eqnarray}&&
	A= {[\kappa\tau_{14}-a_r-2c_r+2(b-c)/r] }/{ (a-b)},
       \label{A=}
\\&& 
	P= 2{[A_r+(A+1/r) ^2] }/{(A+2/r)},
\label{P=}
\\&& 
	Q=\frac{\kappa(\tau_{44}-\lambda)+1/r^2
	-ab-2ac-2bc-c^2 }{A/2+1/r}.\ \ \ \ \ \ \ 
\label{Q=}
\end{eqnarray}
On the other hand, if $a=b$, the functions $c$ and $\phi$ are arbitrary, 
	and the solution is given by
\begin{eqnarray}&&
	a=b=-B-2c,\ \ \ 
       \label{a=}
\\&& 
	2\psi=-\int_r^\infty \tilde Pdr
	-\ln\left[\tilde \eta-\int_r^\infty \tilde Q 
	e^{-\int_r^\infty \tilde Pdr}dr
	\right],\ \ \ \ \ \ 
\label{psi=1}
\end{eqnarray}
with an arbitrary constant $\tilde \eta$ and the functions 
\begin{eqnarray}&&
	B= r\int_r^\infty (\kappa\tau_{14}/r-3c/r^2)dr,
       \label{B=}
\\&& 
	\tilde P= 2{[\phi_{rr}+(\phi_r +1/r) ^2] }/{(\phi_r +2/r)},
\label{P'=}
\\&& 
	\tilde Q= 2[\kappa(\tau_{44}-\lambda)+1/r^2
	+3c^2-B^2]/{(\phi_r +2/r)}.\ \ \ \ \ \ \ 
\label{Q'=}
\end{eqnarray}
The ranges of integration in (\ref{phi=}), (\ref{psi=}),
	(\ref{psi=1}) and (\ref{B=})
	are chosen in accordance with the asymptotic flatness
	of the brane. 
The denominators in (\ref{P=}), (\ref{Q=}), (\ref{P'=}) and (\ref{Q'=})
	are not identically vanishing because, if so, we have 
\begin{eqnarray}&&
	\phi=-2\ln C r
\end{eqnarray}
	with a constant $C$,
	and the brane cannot be asymptotically flat. 
Thus, the solutions for $f$, $h$ and $k$ 
	with $\phi$, $\psi$, $a$, $b$ and $c$ 
	by (\ref{phi=}) and (\ref{psi=}) or (\ref{a=}) and (\ref{psi=1})  
	expire all the solutions of the bulk Einstein equation (\ref{BE}) (alone)
	under the ansatze (a)staticity, (b)sphericity and (c)asymptotic flatness. 
They are not the final solutions 
	because the $T_{IJ}$'s still depend on $f$, $h$ and $k$
	through the matter equations. 
To get final ones, we should solve the coupled system of the equations.
If we add, however, the ansatz (d) of emptiness i.\ e.\ $T_{IJ}=0$,
	the present solutions become final ones.
Thus we have obtained the general solution of 
	the bulk Einstein equation (alone) in the region including the brane 
	under the premised ansatze (a)--(d).

So far we considered the solutions starting with choosing $a$, $b$ and $c$
	as the three arbitrary functions.
We can choose other sets of the arbitrary functions.
If we choose $f$, $h$ and $a+2c$ as arbitrary, 
	the equations (\ref{F140=0}) and (\ref{F440=0}) become, respectively, 
	a linear and a quadratic algebraic equations for $a$ and $b$,
	and they are solved easily.
Then, we can obtain another expression for the general solution 
	of the bulk Einstein equation (alone).

Now we turn to the solution for the ``braneworld".
It is a thin physical object accompanied by 
	matter distribution in the region $|z|<\delta$, 
	where $\delta$ is the infinitesimal thickness of the brane.
The bulk region outside the brane is separated into two regions 
	$D^\pm$ with $\pm z>\delta$.
The Einstein equation inside the brane and 
	the infinitesimal nature of the thickness $\delta$ lead 
	to the Israel junction condition \cite{Israel}.
The extrinsic curvature diag($a,b,c,c$) should have a gap across the brane.
We denote in $\alpha^\pm=\alpha|_{z=\pm \delta}$ for $\alpha=a,b,c$, and 
	$\bar \alpha=(\alpha^++\alpha^-)/2$ and 
	$\Delta \alpha=\alpha^+-\alpha^-$.
The matters obey the equations of motion 
	derived from the action $S_Y+S_{\rm m}$,             
	where the collective mode $Y^I$ is assumed to be separated
	and to obey the Nambu-Goto equation (\ref{NG}).
Here we assume dominance of the collective mode
	in the energy-momentum tensor.
Then, the Nambu-Goto equation and the the Israel junction condition imply
\begin{eqnarray}&&
	\bar a+\bar b+2\bar c=0,
\label{a+b+2c=0}
\\&& 
	\Delta a=\Delta b=\Delta c=\Delta, \ \  \ \ \ \ \ 
\label{Delta}
\end{eqnarray}
where $\Delta$ is a constant.
Since $T_{IJ}^\pm=0$ in $D^\pm$, 
	eqs.\ (\ref{F140=0}) and (\ref{F440=0}) imply
\begin{eqnarray}&&
	\bar a_r+2\bar c_r+ (\bar a-\bar b)\phi_r +2(\bar c-\bar b)/r=0,
\label{barF140=0}
\\&& 
	-\tilde R/2+\bar a\bar b+2\bar a\bar c +2\bar b\bar c +\bar c^2
	+3\Delta^2/2+\kappa\lambda=0. \ \  \ \ \ \ \ 
\label{barF440=0}
\end{eqnarray}
Now, we have three equations (\ref{a+b+2c=0}),
	(\ref{barF140=0}) and (\ref{barF440=0})
	for five functions $\phi$, $\psi$, $\bar a$, $\bar b$ and $\bar c$.
Therefore, the solution includes two arbitrary functions.
If we choose $\bar a$ and $\bar b$ arbitrarily,
	we have the same solution as (\ref{phi=})--(\ref{Q'=})
	with $a$, $b$, $c$, $\tau_{14}$ and $\tau_{44}$
	replaced by $\bar a$, $\bar b$, $-(\bar a+\bar b)/2$, 0
	and $-3\Delta^2/2\kappa$, respectively.
Then, the general solution of the braneworld 
	with the ansatze (a)--(d) is given by
	the recursion formulae (\ref{fn})--(\ref{kn}) with $T_{IJ}=0$.
If we choose $\phi$ and $\psi$ as another choice of the two arbitrary functions, 
	(\ref{barF140=0}) becomes a linear differential equation for $a+2c$
	with the algebraic constraints from (\ref{a+b+2c=0}) and (\ref{barF440=0}).
They give a non-linear differential equation. 
It is expected to have solutions under a certain wide class 
	of the conditions for existence.
Then, we have another form of the general solution of the braneworld 
	with the ansatze (a)--(d).

Therefore the general solution involve large arbitrariness
	which reduces the predictability of the theories.
It includes the cases where
	it coincides with the Schwarzschild solution on the brane.
It also includes, however, continuously deformed solutions 
	which do not satisfy the brane Einstein equation.
It involves arbitrarily deformed Newtonian potentials,
	and arbitrary amount of light deflections and
	planetary perihelion precessions due to solar gravity.
We need further physical prescriptions to make these predictions.
In some models, people require Z2 symmetry with respect to the brane.
In this case, we have
\begin{eqnarray}&&
	a^\pm= b^\pm= c^\pm=\pm \Delta/2, \ \  \ \ \ \ \ 
\label{Z2a+-}
\\&& 
	-\tilde R+3\Delta^2+2\kappa\lambda=0. \ \  \ \ \ \ \ 
\label{Z2Rbr}
\end{eqnarray}
Accordingly $\bar a=\bar b=\bar c=0$, and the arbitrariness is much reduced.
Eq.\ (\ref{Z2Rbr}), however, still leaves an arbitrary function 
	among $\phi$ and $\psi$.
It is the general solution found by Visser and Wiltshire
	\cite{Visser:2002vg}.
Thus, we need further conditions.
For, example, conditions for the bulk behavior may improve the situation.
It is an urgent open problem to be investigated in the future.
Another interesting way out is the idea of the brane induced gravity
	\cite {Akama82}, \cite {Akama87}, \cite{AH},
	\cite {BIG}, \cite {DvaliGabadadze}, \cite {Akama06}, 
	\cite{Gabadadze:2007dv}.
In this case, the Einstein-Hilbert action of the brane 
	is induced through the quantum effects of the matters on the brane,
	and the Einstein gravity emerges effectively
	\cite{InducedGravity}.
The composite metric serves as the effective dynamical variable,
	unlike in the classical braneworld given by (\ref{action}).
Such roles of the quantum-induced composite field 
	have been well understood in some models \cite {ig}.
It provides a promising mechanism for induced field theories.


We would like to thank 
Professor I.~Antoniadis, 
Professor T.~Asaka, 
Professor E.~J.~Copeland, 
Professor G.~R.~Dvali,
Professor G.~Gabadadze,
Professor G.~W.~Gibbons, 
Professor M.~Giovannini,
Professor R.~Gregory, 
Professor K.~Hashimoto, 
Professor A.~Hebecker, 
Professor T.~Inami, 
Professor P.~Kanti, 
Professor A.~Kobakhidze,
Professor S.~Nam,  
Professor I.~P.~Neupane,
Professor I.~Oda, 
Professor M.~Peloso,
Professor S.~Randjbar-Daemi, 
Professor M.~E.~Shaposhnikov, 
Professor M.~Shifman, 
Professor A.~Vainshtein, 
Professor R.~R.~Volkas, 
Professor D.~Wands, 
Professor C.~Wetterich, and 
Professor D.~L.~Wiltshire 
for invaluable discussions and for their kind hospitalities over us
when we visited them.

This work was supported by Grant-in-Aid for Scientific Research,
No.\ 13640297, 17500601, and 22500819
from Japanese Ministry of Education, Culture, Sports, Science and Technology.

\end{document}